# Event-by-event correlations between $\Lambda$ ($\bar{\Lambda}$) hyperon global polarization and handedness with charged hadron azimuthal separation in Au+Au collisions at $\sqrt{s_{\mathrm{NN}}} = 27$ GeV from STAR

M. I. Abdulhamid,[4] B. E. Aboona,[54] J. Adam,[15] J. R. Adams,[39] G. Agakishiev,[29] I. Aggarwal,[40] M. M. Aggarwal,[40] Z. Ahammed,[60] A. Aitbaev,[29] I. Alekseev,[2,36] D. M. Anderson,[54] A. Aparin,[29] S. Aslam,[25] J. Atchison,[1] G. S. Averichev,[29] V. Bairathi,[52] W. Baker,[11] J. G. Ball Cap,[21] K. Barish,[11] P. Bhagat,[28] A. Bhasin,[28] S. Bhatta,[51] I. G. Bordyuzhin,[2] J. D. Brandenburg,[39] A. V. Brandin,[36] X. Z. Cai,[49] H. Caines,[62] M. Calderón de la Barca Sánchez,[9] D. Cebra,[9] J. Ceska,[15] I. Chakaberia,[32] B. K. Chan,[10] Z. Chang,[26] A. Chatterjee,[16] D. Chen,[11] J. Chen,[48] J. H. Chen,[19] Z. Chen,[48] J. Cheng,[56] Y. Cheng,[10] S. Choudhury,[19] W. Christie,[6] X. Chu,[6] H. J. Crawford,[8] G. Dale-Gau,[13] A. Das,[15] M. Daugherity,[1] T. G. Dedovich,[29] I. M. Deppner,[20] A. A. Derevschikov,[41] A. Dhamija,[40] L. Di Carlo,[61] L. Didenko,[6] P. Dixit,[23] X. Dong,[32] J. L. Drachenberg,[1] E. Duckworth,[30] J. C. Dunlop,[6] J. Engelage,[8] G. Eppley,[43] S. Esumi,[57] O. Evdokimov,[13] A. Ewigleben,[33] O. Eyser,[6] R. Fatemi,[31] S. Fazio,[7] C. J. Feng,[38] Y. Feng,[42] E. Finch,[50] Y. Fisyak,[6] F. A. Flor,[62] C. Fu,[27] F. Geurts,[43] N. Ghimire,[53] A. Gibson,[59] K. Gopal,[24] X. Gou,[48] D. Grosnick,[59] A. Gupta,[28] A. Hamed,[4] Y. Han,[43] M. D. Harasty,[9] J. W. Harris,[62] H. Harrison-Smith,[31] W. He,[19] X. H. He,[27] Y. He,[48] C. Hu,[27] Q. Hu,[27] Y. Hu,[32] H. Huang,[38] H. Z. Huang,[10] S. L. Huang,[51] T. Huang,[13] X. Huang,[56] Y. Huang,[56] Y. Huang,[12] T. J. Humanic,[39] D. Isenhower,[1] M. Isshiki,[57] W. W. Jacobs,[26] A. Jalotra,[28] C. Jena,[24] Y. Ji,[32] J. Jia,[6,51] C. Jin,[43] X. Ju,[45] E. G. Judd,[8] S. Kabana,[52] M. L. Kabir,[11] D. Kalinkin,[31] K. Kang,[56] D. Kapukchyan,[11] D. Keane,[30] A. Kechechyan,[29] M. Kelsey,[61] B. Kimelman,[9] A. Kiselev,[6] A. G. Knospe,[33] H. S. Ko,[32] L. Kochenda,[36] A. A. Korobitsin,[29] P. Kravtsov,[36] L. Kumar,[40] S. Kumar,[27] R. Kunnawalkam Elayavalli,[62] R. Lacey,[51] J. M. Landgraf,[6] A. Lebedev,[6] R. Lednicky,[29] J. H. Lee,[6] Y. H. Leung,[20] N. Lewis,[6] C. Li,[48] W. Li,[43] X. Li,[45] Y. Li,[45] Y. Li,[56] Z. Li,[45] X. Liang,[11] Y. Liang,[30] T. Lin,[48] C. Liu,[27] F. Liu,[12] G. Liu,[46] H. Liu,[26] H. Liu,[12] L. Liu,[12] T. Liu,[62] X. Liu,[39] Y. Liu,[54] Z. Liu,[12] T. Ljubicic,[6] W. J. Llope,[61] O. Lomicky,[15] R. S. Longacre,[6] E. M. Loyd,[11] T. Lu,[27] N. S. Lukow,[53] X. F. Luo,[12] V. B. Luong,[29] L. Ma,[19] R. Ma,[6] Y. G. Ma,[19] N. Magdy,[51] D. Mallick,[37] S. Margetis,[30] H. S. Matis,[32] J. A. Mazer,[44] G. McNamara,[61] K. Mi,[12] N. G. Minaev,[41] B. Mohanty,[37] M. M. Mondal,[37] I. Mooney,[62] D. A. Morozov,[41] A. Mudrokh,[29] M. I. Nagy,[17] A. S. Nain,[40] J. D. Nam,[53] M. Nasim,[23] D. Neff,[10] J. M. Nelson,[8] D. B. Nemes,[62] M. Nie,[48] G. Nigmatkulov,[36] T. Niida,[57] R. Nishitani,[57] L. V. Nogach,[41] T. Nonaka,[57] G. Odyniec,[32] A. Ogawa,[6] S. Oh,[47] V. A. Okorokov,[36] K. Okubo,[57] B. S. Page,[6] R. Pak,[6] J. Pan,[54] A. Pandav,[37] A. K. Pandey,[27] Y. Panebratsev,[29] T. Pani,[44] P. Parfenov,[36] A. Paul,[11] C. Perkins,[8] B. R. Pokhrel,[53] M. Posik,[53] T. Protzman,[33] N. K. Pruthi,[40] J. Putschke,[61] Z. Qin,[56] H. Qiu,[27] A. Quintero,[53] C. Racz,[11] S. K. Radhakrishnan,[30] N. Raha,[61] R. L. Ray,[55] H. G. Ritter,[32] C. W. Robertson,[42] O. V. Rogachevsky,[29] M. A. Rosales Aguilar,[31] D. Roy,[44] L. Ruan,[6] A. K. Sahoo,[23] N. R. Sahoo,[48] H. Sako,[57] S. Salur,[44] E. Samigullin,[2] S. Sato,[57] W. B. Schmidke,[6] N. Schmitz,[34] J. Seger,[14] R. Seto,[11] P. Seyboth,[34] N. Shah,[25] E. Shahaliev,[29] P. V. Shanmuganathan,[6] T. Shao,[19] M. Sharma,[28] N. Sharma,[23] R. Sharma,[24] S. R. Sharma,[24] A. I. Sheikh,[30] D. Y. Shen,[19] K. Shen,[45] S. S. Shi,[12] Y. Shi,[48] Q. Y. Shou,[19] F. Si,[45] J. Singh,[40] S. Singha,[27] P. Sinha,[24] M. J. Skoby,[5,42] Y. Söhngen,[20] Y. Song,[62] B. Srivastava,[42] T. D. S. Stanislaus,[59] D. J. Stewart,[61] M. Strikhanov,[36] B. Stringfellow,[42] Y. Su,[45] C. Sun,[51] X. Sun,[27] Y. Sun,[45] Y. Sun,[22] B. Surrow,[53] D. N. Svirida,[2] Z. W. Sweger,[9] A. Tamis,[62] A. H. Tang,[6] Z. Tang,[45] A. Taranenko,[36] T. Tarnowsky,[35] J. H. Thomas,[32] D. Tlusty,[14] T. Todoroki,[57] M. V. Tokarev,[29] C. A. Tomkiel,[33] S. Trentalange,[10] R. E. Tribble,[54] P. Tribedy,[6] O. D. Tsai,[10,6] C. Y. Tsang,[30,6] Z. Tu,[6] J. Tyler,[54] T. Ullrich,[6] D. G. Underwood,[3,59] I. Upsal,[45] G. Van Buren,[6] A. N. Vasiliev,[41,36] V. Verkest,[61] F. Videbæk,[6] S. Vokal,[29] S. A. Voloshin,[61] F. Wang,[42] G. Wang,[10] J. S. Wang,[22] X. Wang,[48] Y. Wang,[45] Y. Wang,[12] Y. Wang,[56] Z. Wang,[48] J. C. Webb,[6] P. C. Weidenkaff,[20] G. D. Westfall,[35] H. Wieman,[32] G. Wilks,[13] S. W. Wissink,[26] J. Wu,[12] J. Wu,[27] X. Wu,[10] Y. Wu,[11] B. Xi,[49] Z. G. Xiao,[56] G. Xie,[58] W. Xie,[42] H. Xu,[22] N. Xu,[32] Q. H. Xu,[48] Y. Xu,[48] Y. Xu,[12] Z. Xu,[6] Z. Xu,[10] G. Yan,[48] Z. Yan,[51] C. Yang,[48] Q. Yang,[48] S. Yang,[46] Y. Yang,[38] Z. Ye,[43] Z. Ye,[13] L. Yi,[48] K. Yip,[6] Y. Yu,[48] W. Zha,[45] C. Zhang,[51] D. Zhang,[12] J. Zhang,[48] S. Zhang,[45] W. Zhang,[46] X. Zhang,[27] Y. Zhang,[27] Y. Zhang,[45] Y. Zhang,[12] Z. J. Zhang,[38] Z. Zhang,[6] Z. Zhang,[13] F. Zhao,[27] J. Zhao,[19] M. Zhao,[6] C. Zhou,[19] J. Zhou,[45] S. Zhou,[12] Y. Zhou,[12] X. Zhu,[56] M. Zurek,[3,6] and M. Zyzak[18]

(STAR Collaboration)

[1]*Abilene Christian University, Abilene, Texas 79699*
[2]*Alikhanov Institute for Theoretical and Experimental Physics NRC "Kurchatov Institute", Moscow 117218*
[3]*Argonne National Laboratory, Argonne, Illinois 60439*
[4]*American University in Cairo, New Cairo 11835, Egypt*
[5]*Ball State University, Muncie, Indiana, 47306*

[6]Brookhaven National Laboratory, Upton, New York 11973
[7]University of Calabria & INFN-Cosenza, Rende 87036, Italy
[8]University of California, Berkeley, California 94720
[9]University of California, Davis, California 95616
[10]University of California, Los Angeles, California 90095
[11]University of California, Riverside, California 92521
[12]Central China Normal University, Wuhan, Hubei 430079
[13]University of Illinois at Chicago, Chicago, Illinois 60607
[14]Creighton University, Omaha, Nebraska 68178
[15]Czech Technical University in Prague, FNSPE, Prague 115 19, Czech Republic
[16]National Institute of Technology Durgapur, Durgapur - 713209, India
[17]ELTE Eötvös Loránd University, Budapest, Hungary H-1117
[18]Frankfurt Institute for Advanced Studies FIAS, Frankfurt 60438, Germany
[19]Fudan University, Shanghai, 200433
[20]University of Heidelberg, Heidelberg 69120, Germany
[21]University of Houston, Houston, Texas 77204
[22]Huzhou University, Huzhou, Zhejiang 313000
[23]Indian Institute of Science Education and Research (IISER), Berhampur 760010 , India
[24]Indian Institute of Science Education and Research (IISER) Tirupati, Tirupati 517507, India
[25]Indian Institute Technology, Patna, Bihar 801106, India
[26]Indiana University, Bloomington, Indiana 47408
[27]Institute of Modern Physics, Chinese Academy of Sciences, Lanzhou, Gansu 730000
[28]University of Jammu, Jammu 180001, India
[29]Joint Institute for Nuclear Research, Dubna 141 980
[30]Kent State University, Kent, Ohio 44242
[31]University of Kentucky, Lexington, Kentucky 40506-0055
[32]Lawrence Berkeley National Laboratory, Berkeley, California 94720
[33]Lehigh University, Bethlehem, Pennsylvania 18015
[34]Max-Planck-Institut für Physik, Munich 80805, Germany
[35]Michigan State University, East Lansing, Michigan 48824
[36]National Research Nuclear University MEPhI, Moscow 115409
[37]National Institute of Science Education and Research, HBNI, Jatni 752050, India
[38]National Cheng Kung University, Tainan 70101
[39]The Ohio State University, Columbus, Ohio 43210
[40]Panjab University, Chandigarh 160014, India
[41]NRC "Kurchatov Institute", Institute of High Energy Physics, Protvino 142281
[42]Purdue University, West Lafayette, Indiana 47907
[43]Rice University, Houston, Texas 77251
[44]Rutgers University, Piscataway, New Jersey 08854
[45]University of Science and Technology of China, Hefei, Anhui 230026
[46]South China Normal University, Guangzhou, Guangdong 510631
[47]Sejong University, Seoul, 05006, South Korea
[48]Shandong University, Qingdao, Shandong 266237
[49]Shanghai Institute of Applied Physics, Chinese Academy of Sciences, Shanghai 201800
[50]Southern Connecticut State University, New Haven, Connecticut 06515
[51]State University of New York, Stony Brook, New York 11794
[52]Instituto de Alta Investigación, Universidad de Tarapacá, Arica 1000000, Chile
[53]Temple University, Philadelphia, Pennsylvania 19122
[54]Texas A&M University, College Station, Texas 77843
[55]University of Texas, Austin, Texas 78712
[56]Tsinghua University, Beijing 100084
[57]University of Tsukuba, Tsukuba, Ibaraki 305-8571, Japan
[58]University of Chinese Academy of Sciences, Beijing, 101408
[59]Valparaiso University, Valparaiso, Indiana 46383
[60]Variable Energy Cyclotron Centre, Kolkata 700064, India
[61]Wayne State University, Detroit, Michigan 48201
[62]Yale University, New Haven, Connecticut 06520

Global polarizations ($P$) of $\Lambda$ ($\bar{\Lambda}$) hyperons have been observed in non-central heavy-ion collisions. The strong magnetic field primarily created by the spectator protons in such collisions would split the $\Lambda$ and $\bar{\Lambda}$ global polarizations ($\Delta P = P_{\Lambda} - P_{\bar{\Lambda}} < 0$). Additionally, quantum chromodynamics (QCD) predicts topological charge fluctuations in vacuum, resulting in a chirality imbalance or parity violation in a local domain. This would give rise to an imbalance ($\Delta n = \frac{N_{\rm L} - N_{\rm R}}{\langle N_{\rm L} + N_{\rm R} \rangle} \neq 0$) between

left- and right-handed $\Lambda$ ($\bar{\Lambda}$) as well as a charge separation along the magnetic field, referred to as the chiral magnetic effect (CME). This charge separation can be characterized by the parity-even azimuthal correlator ($\Delta\gamma$) and parity-odd azimuthal harmonic observable ($\Delta a_1$). Measurements of $\Delta P$, $\Delta\gamma$, and $\Delta a_1$ have not led to definitive conclusions concerning the CME or the magnetic field, and $\Delta n$ has not been measured previously. Correlations among these observables may reveal new insights. This paper reports measurements of correlation between $\Delta n$ and $\Delta a_1$, which is sensitive to chirality fluctuations, and correlation between $\Delta P$ and $\Delta\gamma$ sensitive to magnetic field in Au+Au collisions at 27 GeV. For both measurements, no correlations have been observed beyond statistical fluctuations.

## 1. INTRODUCTION

In non-central heavy-ion collisions, due to finite impact parameter, only a fraction of nucleons (called participants) participate in the collision, while the others (called spectators) are out of the collision zone and continue along the beam lines. The spectator protons are predicted to create, in the first moments of the collision, a magnetic field [1, 2] that is strong enough to align quark spin either parallel or anti-parallel to the magnetic field, depending on the quark electric charge. The positively and negatively charged quarks of the same chirality would thus have opposite momentum directions along the magnetic field. This would result in a charge separation if the numbers of left- and right-handed quarks are imbalanced, a phenomenon called the chiral magnetic effect (CME) [1, 2]. Such a chirality imbalance has indeed been predicted to occur because of the chiral anomaly in Quantum Chromodynamics (QCD) [3]. It is a direct result of quark interactions with gluon fields possessing, due to fluctuations, non-zero topological charges ($Q_w$). Such gluon field domains explicitly break the parity ($\mathcal{P}$) and charge-parity ($\mathcal{CP}$) symmetry and are a fundamental ingredient of QCD [1, 3–5].

The azimuthal distribution of particles in each event can be expanded into Fourier series:

$$\frac{2\pi}{N^{\pm}}\frac{\mathrm{d}N^{\pm}}{\mathrm{d}\phi} = 1 + 2a_1^{\pm}\sin(\phi^{\pm} - \Psi_{\mathrm{RP}}) + \sum_{n=1}^{+\infty} 2v_n \cos n(\phi^{\pm} - \Psi_{\mathrm{RP}}), \tag{1}$$

where the superscripts $\pm$ indicate the charge sign; $\phi$ represents the azimuthal angle of particles. The reaction plane (RP) is spanned by the beam direction and the impact parameter, and its azimuthal angle is denoted by $\Psi_{\mathrm{RP}}$. Based on Eq. 1, many observables are proposed to measure the CME, like the parity-odd $\Delta a_1$ variable [1, 6] (Sec. 2.3), and the parity-even $\Delta\gamma$ variable [7] (Sec. 2.4). The parity-odd $\Delta a_1$ observable vanishes in event average because of the random fluctuations of topological charges. Experiments have focused on the parity-even $\Delta\gamma$ correlator observable. So far, no definitive conclusion on the CME has been reached by $\Delta\gamma$ measurements at RHIC in Au+Au [6, 8–11] and $d$+Au [12] collisions or at the LHC in Pb+Pb [13–17] and $p$+Pb [14, 15]) collisions. The main difficulty in the $\Delta\gamma$ interpretation is background contamination arising from particle correlations coupled with elliptic flow [7, 18–23]. Many methods have been proposed to reduce or remove the backgrounds [10, 15, 16, 24–27] but with limited success.

The chirality preference of quarks in the collision zone can be inherited by $\Lambda$ hyperons in the final state [28]. In this paper, $\Lambda$ denotes both $\Lambda$ and $\bar{\Lambda}$ except otherwise specified. $\Lambda$ hyperons can be detected in experiments via their main decay channel $\Lambda \to p + \pi^-$ [29, 30]. Their handedness (the sign of helicity) can be measured by their decay topology (Sec. 2.5). In each event, the normalized handedness imbalance $\Delta n$ can be defined from the measured numbers of left-handed and right-handed $\Lambda$'s (Sec. 2.5). Similar to $\Delta a_1$, $\Delta n$ is parity-odd, therefore its average over many events must be zero [6]. Although vanishing trivially in their event averages, $\Delta a_1$ and $\Delta n$ both come from the same chirality anomaly in each event, so their event-by-event correlation could be non-zero [28]. For example, if the topological charge is negative ($Q_w < 0$), then the $\Delta n$ values of $u$, $d$, and $s$ quark would all be negative [28, 31]. The $\Lambda$ hyperon would then be expected to inherit the finite $\Delta n$ from the $s$ quark [32–34]. Meanwhile, the negative $\Delta n$ values of $u$ and $d$ quarks would result in a positive $\Delta a_1$. Similarly, $Q_w > 0$ would yield positive $\Delta n$ and negative $\Delta a_1$. Therefore, the quantum chiral anomaly would result in a negative correlation between $\Delta n$ and $\Delta a_1$. We note that $\Lambda$ hyperons contain both produced and transported quarks, which may be affected differently by the topological domain. However, the sign of their contributions are expected to be the same, so the discussion above should still be valid. We also note that the $s$ quark has finite mass, larger than $u$ and $d$, so its chirality might flip during its evolution and interaction with the environment [28, 35]. If so, the final-state $\Lambda$ may reflect only part of the initial-state chirality imbalance.

Besides a positive $\Delta\gamma$ signal from the CME [7], the magnetic field can have another consequence, namely a difference in the $\Lambda$ and $\bar{\Lambda}$ global polarizations. These global polarizations are mainly caused by the vorticity arising from the total angular momentum of the collision participants, which are equal for $\Lambda$ and $\bar{\Lambda}$ [36–42]. However, the magnetic field, aligned on average with the total angular momentum, can cause a difference in polarization between the two species due to their opposite magnetic moments [29]; it enhances the polarization of $\bar{\Lambda}$ and reduces that of $\Lambda$ [43]. Thus, the polarization dif-ference between $\Lambda$ ($P_\Lambda$) and $\bar{\Lambda}$ ($P_{\bar{\Lambda}}$), $\Delta P = P_\Lambda - P_{\bar{\Lambda}}$, has been proposed as a probe of the magnetic field. Current statistical precision has not allowed a firm conclusion [39, 42]. Large fluctuations in the magnetic field have been predicted [43–45], so correlations between $\Delta P$ and $\Delta\gamma$ may be more sensitive than individual measurements of the averages. Since the magnetic field yields a positive $\Delta\gamma$ and a negative $\Delta P$, a negative correlation would be a strong indication of the presence of magnetic field. We note that some of the final-state $\Lambda$'s come from the decay of heavier particles like $\Sigma$, $\Xi$, $\Omega$, and this feed-down effect can dilute the $\Lambda$ handedness and polarization measurements [46]. Since those heavier particles are also subjected to the same physics–the chirality anomaly, vorticity, and magnetic field, including those feed-down $\Lambda$'s should not change the qualitative expectation for our correlation measurements.

This paper reports measurements of event-by-event correlations between $\Delta n$ and $\Delta a_1$ and between $\Delta P$ and $\Delta\gamma$ in Au+Au collisions at $\sqrt{s_{\rm NN}} = 27$ GeV from STAR Run18 data. The rest of the paper is organized as follows: Section 2 presents the definitions of the observables used in this study and describes the methodologies of their measurements with analysis details. Section 3 discusses the systematic uncertainties in our measurements. Section 4 reports our results. Section 5 summarizes the paper.

## 2. EXPERIMENT AND DATA ANALYSIS

The Au+Au collision data at $\sqrt{s_{\rm NN}} = 27$ GeV were taken in 2018 by the STAR experiment, with the newly installed Event Plane Detector (EPD) [47] covering the pseudorapidity range $2.1 < |\eta| < 5.1$ [47]. Events with the minimum-bias trigger are used for this analysis. In each event, the primary vertex measured with the Time Projection Chamber (TPC) [48, 49] is required to have $V_r = \sqrt{V_x^2 + V_y^2} < 2$ cm and $|V_z| < 70$ cm, and its longitudinal distance from the Vertex Position Detector (VPD) [50] measurement is required to satisfy $|V_z^{\rm VPD} - V_z| < 3$ cm, where $z$ is the beam direction and $r$ stands for the transverse direction perpendicular to $z$. After those event-level selections, there are about 400 million events left. For TPC tracking quality, the number of hits for track fitting is required to be no less than 15 for all the detected particles. This study uses centrality defined by the measured particle multiplicity in $|\eta| < 0.5$ [51].

### 2.1. Event plane reconstruction

In experiments, the reaction plane is unknown, but can be estimated by the event planes reconstructed using various detectors, each of which has its own resolution for finding the reaction plane. This study uses the EPD [47] to get the first- and second-order event planes. With respect to reaction plane ($\Psi_{\rm RP}$), the resolution $R_{nk}$ is defined as

$$R_{nk} \equiv \langle \cos(k(\Psi_n - \Psi_{\rm RP}))\rangle. \quad (2)$$

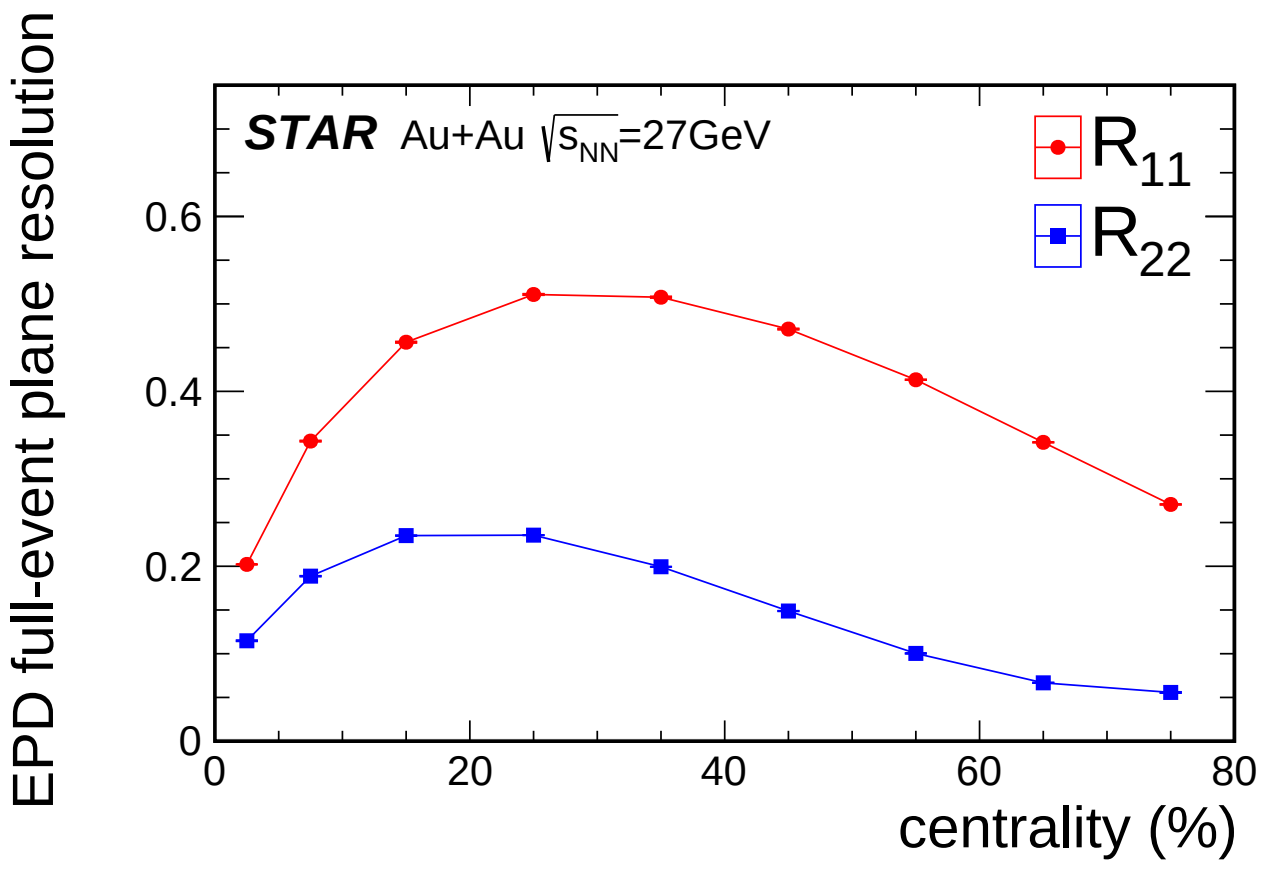

FIG. 1. The resolutions as functions of centrality for full EPD ($2.1 < |\eta| < 5.1$, including both sides) event planes ($R_{11}$ for $\Psi_1$, $R_{22}$ for $\Psi_2$) in Au+Au at $\sqrt{s_{\rm NN}} = 27$ GeV. The statistical and systematic uncertainties are too small to be visible.

For $k = n$, subevent $R_{nn}^{\rm sub}$ can be estimated by

$$R_{nn}^{\rm sub} = \sqrt{\langle \cos(n(\Psi_n^E - \Psi_n^W))\rangle}, \quad (3)$$

where $\Psi_n^E$ ($\Psi_n^W$) is the $n^{\rm th}$-order event plane reconstructed from East (West) EPD. The full resolution $R_{nn}$ and also $R_{nk}$ ($k \neq n$) can be then obtained by using Bessel functions [52]. In this study, the full EPD event planes $\Psi_1$, $\Psi_2$ and their corresponding resolutions $R_{11}$, $R_{22}$ are used (Fig. 1). The resolution correction is of event average, not on an event-by-event basis. Its purpose is to correct the overall magnitude of the quantities calculated w.r.t. EP, such as $\Delta a_1$, $\Delta\gamma$, $\Lambda$ polarizations, and the correlations.

### 2.2. $\Lambda$ and $\bar{\Lambda}$ Reconstruction

The Kalman filter method and `KFParticle` package [53, 54] are used to reconstruct $\Lambda$ from the decay $\Lambda \to p + \pi^-$ (again the notation includes $\bar{\Lambda} \to \bar{p} + \pi^+$ except otherwise noted) [29, 30]. The final-state hadrons ($p$, $\bar{p}$, $\pi^+$, $\pi^-$) are identified by the TPC and TOF (Time Of Flight) [50] detectors. To optimize the statistics, no other kinematic selection is applied to those hadrons. Instead, the reconstructed $\Lambda$ is selected with a transverse momentum requirement of 0.4 GeV/$c < p_T <$ 3.0 GeV/$c$.

The mass spectra of the reconstructed $\Lambda$ and $\bar{\Lambda}$ are shown in Fig. 2 (a), while Fig. 2 (b) shows the peak region ($m_\Lambda \pm 0.005$ GeV/$c^2$, bounded by red dashed lines) and the off-peak regions (1.090–1.105 GeV/$c^2$ and 1.125–1.180 GeV/$c^2$ bounded by blue dashed lines). The peak region is a mixture of signal and background, so the mass spectra of this region are fitted by a function including signal (double-Gaussian) and background (1st-order polynomial). Then, the number of signal particles ($S$) and background particles ($B$) can be extracted in each centrality class. The $S/B$ ratio is shown in Fig. 2 (c). For further purity correction, the off-peak regions (Fig. 2 (b)) are used to estimate the background baseline. The sharp $\Lambda$ peak and large $S/B$ ratio shown in Fig. 2 indicate the high efficiency of the `KFParticle` package for $\Lambda$ reconstruction.

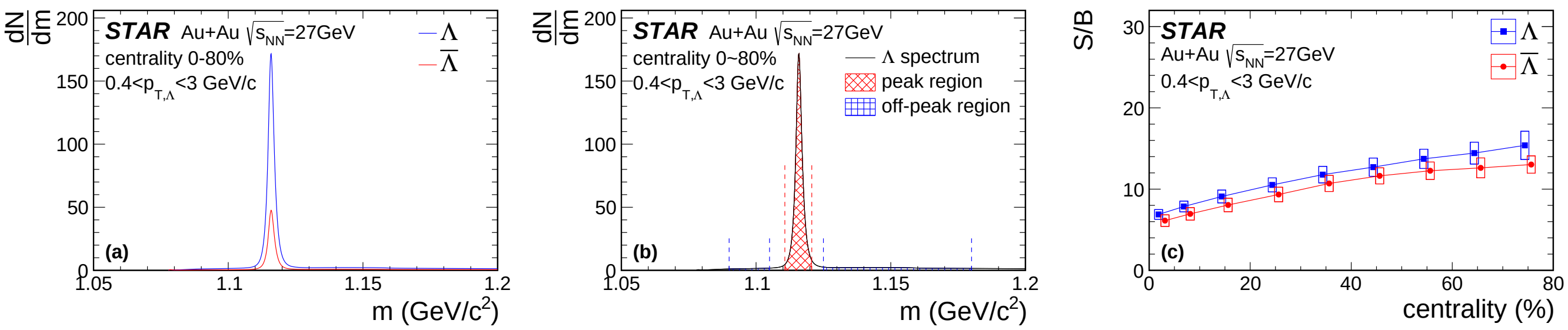

FIG. 2. (a) The invariant mass spectra of the reconstructed $\Lambda$ with $0.4$ GeV$/c < p_T < 3.0$ GeV$/c$ in minimum-bias (centrality range 0–80%) Au+Au collisions at $\sqrt{s_{\rm NN}} = 27$ GeV. (b) Illustration of mass regions for $\Lambda$ candidates and combinatoric background. The candidates are selected in the peak region ($m_\Lambda \pm 0.005$ GeV$/c^2$ bounded by red dashed lines) and the background is assessed by the off-peak region (1.090–1.105 GeV$/c^2$ and 1.125–1.180 GeV$/c^2$ bounded by blue dashed lines). (c) Signal to background ratio in the $\Lambda$ mass on-peak region as functions of centrality. The statistical uncertainty is too small to be visible, while the systematic uncertainty is shown by hollow boxes. The $\Lambda$ data points are shifted slightly to the left along the $x$-axis, while $\bar{\Lambda}$ to the right symmetrically, for better visibility.

### 2.3. Charge Separation $\Delta a_1$ of Unidentified Charged Hadrons

In each event, the coefficients $a_1^\pm$ and $\Delta a_1$ are calculated from unidentified charged hadrons as follows:

$$\begin{aligned} a_1^+ =&\langle \sin(\phi^+ - \Psi_{\rm RP})\rangle = \langle \sin(\phi^+ - \Psi_1)\rangle / R_{11}, \\ a_1^- =&\langle \sin(\phi^- - \Psi_{\rm RP})\rangle = \langle \sin(\phi^- - \Psi_1)\rangle / R_{11}, \\ \Delta a_1 =& a_1^+ - a_1^- , \end{aligned} \tag{4}$$

where the superscripts "±" indicate the charge sign of the particle. The EPD $\Psi_1$ is used to estimate RP, with corresponding resolution correction $R_{11}$. As a parity-odd observable, $\Delta a_1$ (also $a_1^+$, $a_1^-$) averages to zero over many events because of random topological charge fluctuations from event to event.

The CME observables ($\Delta a_1$, $\Delta\gamma$) are calculated using the unidentified charged hadrons with selections $-1 \leq \eta \leq 1$, $0.2$ GeV$/c \leq p_T \leq 2.0$ GeV$/c$. The number of TPC fit points on the particle track must be larger than or equal to 15, and larger than 0.52 times the maximum number of fit points for a given track trajectory to avoid split tracks. To focus on the primary particles, the distance of the closest approach to the collision primary vertex (DCA) is required to be smaller than 1 cm. When forming correlation with on-peak (off-peak) $\Lambda$ handedness imbalance, $\Delta a_1$ is calculated without the decay daughters from $\Lambda$ candidates from the on-peak (off-peak) region.

### 2.4. Correlator $\Delta\gamma$ of Unidentified Charged Hadrons

An EP-dependent correlator $\Delta\gamma$ ($\equiv \gamma_{\rm OS} - \gamma_{\rm SS}$) [7] is widely used in CME studies. $\Delta\gamma$ is calculated using the same unidentified charged hadrons that are used for $\Delta a_1$. The definitions of $\gamma_{\rm OS}$ and $\gamma_{\rm SS}$ are as follows:

$$\begin{aligned} \gamma_{\rm OS} =&\langle \cos(\phi_\alpha^\pm + \phi_\beta^\mp - 2\Psi_{\rm RP})\rangle \\ =&\langle \cos(\phi_\alpha^\pm + \phi_\beta^\mp - 2\Psi_2)\rangle / R_{22}, \\ \gamma_{\rm SS} =&\langle \cos(\phi_\alpha^\pm + \phi_\beta^\pm - 2\Psi_{\rm RP})\rangle \\ =&\langle \cos(\phi_\alpha^\pm + \phi_\beta^\pm - 2\Psi_2)\rangle / R_{22}, \\ \Delta\gamma =&\gamma_{\rm OS} - \gamma_{\rm SS}, \end{aligned} \tag{5}$$

where the subscripts $\alpha$ and $\beta$ denote two different (primary) particles in the same event, OS (SS) stands for opposite-sign (same-sign) pairs. To subtract the charge-independent background contributions (e.g., momentum conservation, inter-jet correlation, etc.), the difference between $\gamma_{\rm OS}$ and $\gamma_{\rm SS}$ is taken. The CME would cause a positive $\Delta\gamma$ signal, and there are backgrounds that can also fake this positive signal [7, 18–23].

Similar to $\Delta a_1$, $\Delta\gamma$ needs to be calculated from the primary particles, so the same DCA selections are applied and the decay daughters from $\Lambda$ are removed. When forming correlation with on-peak (off-peak) $\Lambda$ polarizations, $\Delta\gamma$ calculation excludes the decay daughters from $\Lambda$ candidates from the on-peak (off-peak) region.

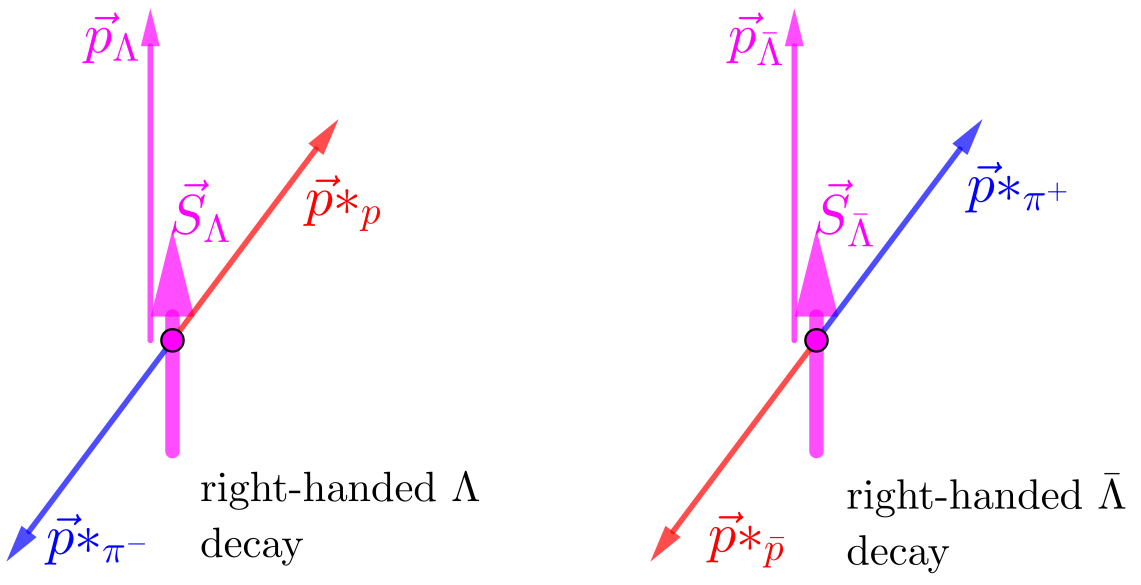

FIG. 3. Schematic diagrams for the decay topology of right-handed Λ (left diagram) and $\bar{\Lambda}$ (right diagram) in their rest frame. Right-handedness is taken as an example. The Λ momentum vector represents the momentum in the laboratory frame.

### 2.5. Λ and $\bar{\Lambda}$ Handedness

For a decay $\Lambda \to p + \pi^-$ in the rest frame of Λ, the decay daughter proton's momentum $\vec{p}_p^*$ tends to distribute around the spin direction of Λ. For $\bar{\Lambda}$, there is a sign difference, i.e., the decay daughter antiproton momentum tends to distribute opposite to the spin direction of $\bar{\Lambda}$. The momentum of each Λ, $\vec{p}_\Lambda$, can be reconstructed from the measured momentum of its decay daughters. Then, the handedness of Λ can be estimated by $\vec{p}_p^* \cdot \vec{p}_\Lambda$.

$$\begin{cases} \vec{p}_p^* \cdot \vec{p}_\Lambda < 0 & \Rightarrow \quad \Lambda_L : \text{“left-handed” } \Lambda \\ \vec{p}_p^* \cdot \vec{p}_\Lambda > 0 & \Rightarrow \quad \Lambda_R : \text{“right-handed” } \Lambda \end{cases}$$
$$\begin{cases} \vec{p}_{\bar{p}}^* \cdot \vec{p}_{\bar{\Lambda}} < 0 & \Rightarrow \quad \bar{\Lambda}_R : \text{“right-handed” } \bar{\Lambda} \\ \vec{p}_{\bar{p}}^* \cdot \vec{p}_{\bar{\Lambda}} > 0 & \Rightarrow \quad \bar{\Lambda}_L : \text{“left-handed” } \bar{\Lambda} \end{cases} \quad (6)$$

Figure 3 shows the schematics for the right-handed Λ and $\bar{\Lambda}$ in their respective rest frame.

In this study, only the observed number of left/right-handed Λ ($N_L^{\rm obs}$, $N_R^{\rm obs}$) is considered, whose difference is referred to as

$$\Delta n^{\rm obs} \equiv \frac{N_L^{\rm obs} - N_R^{\rm obs}}{\langle N_L^{\rm obs} + N_R^{\rm obs} \rangle}. \quad (7)$$

The superscript “obs” stands for the “observed” handedness. The denominator is the event average of the measured number of Λ in a given centrality class. The value of $\Delta n^{\rm obs}$ will be calculated for Λ, $\bar{\Lambda}$, and their sum respectively. Figure 4 shows the mass distribution and $S/B$ ratio of the peak region for Λ with measured handedness.

Figure 5 shows the event average of $N^{\rm obs}$ for left- and right-handed Λ in each centrality class, without correction for the Λ reconstruction inefficiency, which is discussed in the next paragraph. The “on-peak total” (green square) is calculated from all Λ candidates in the peak region. The “off-peak bkg” (blue circle) is calculated from all Λ candidates in the off-peak regions. The “on-peak signal” (red triangle) is calculated from “on-peak total” with the corresponding $S/B$ ratio, separately for left/right-handed $\Lambda/\bar{\Lambda}$ (see Fig. 4 (c)). The average Λ numbers per event are less than 2 in central collisions and down to $10^{-3}$ in peripheral collisions, much smaller than the charged particle multiplicity, so the exclusion of those decay daughters from CME observables should not make a visible difference.

Averaged over many events, the true handedness of Λ must be just as often left as right (within statistical precision) due to global parity conservation. However, detector effects can bias the observed handedness in one direction. In a given event, the numbers of true left-handed and right-handed Λ, $N_L$ and $N_R$, can be affected differently by the topological charge fluctuations. However, the event averages, $\langle N_L \rangle$ and $\langle N_R \rangle$, should be the same within statistical precision, because the topological charge fluctuations are totally random from event to event. Nevertheless, Fig. 5 shows $\langle N_L^{\rm obs}(\Lambda) \rangle \gg \langle N_R^{\rm obs}(\Lambda) \rangle$ and $\langle N_L^{\rm obs}(\bar{\Lambda}) \rangle \ll \langle N_R^{\rm obs}(\bar{\Lambda}) \rangle$. This asymmetry results from the Λ reconstruction inefficiency [42, 55]. This comes from the combination of two effects: First, different handedness results in very different distributions of $p_T$ for the daughter pions. Second, the STAR TPC tracking efficiency becomes lower with decreasing particle $p_T$. Thus, detection efficiencies are very different for left- and right-handed Λ. Figure 6 illustrates the low and high efficiency cases for the decay $\Lambda \to p + \pi^-$. If the decay daughter $\pi^-$'s momentum in the Λ-rest frame is opposite to the decay parent Λ's momentum in the lab frame (observed right-handed, cf. Eq. 6), then, after Lorentz boost, the momentum of that $\pi^-$ in the lab frame would be relatively small, so that the TPC would have a lower efficiency to detect this low-$p_T$ Λ. By contrast, if the decay daughter $\pi^-$ is in the same direction as the decay parent Λ (observed left-handed, cf. Eq. 6), the detector efficiency is relatively high. As the decay daughter proton's momentum is also used in the Λ-rest frame to estimate the Λ handedness, more left(right)-handed $\Lambda(\bar{\Lambda})$ decays are measured by TPC due to this detector inefficiency.

### 2.6. Λ and $\bar{\Lambda}$ Global Polarization

The polarization of Λ can be measured [38] from the distribution of decay daughter protons with respect to the event plane.

$$P_\Lambda = \frac{-8}{\pi \alpha_\Lambda} \langle \sin(\phi_p^* - \Psi_{\rm RP}) \rangle = \frac{-8}{\pi \alpha_\Lambda R_{11}} \langle \sin(\phi_p^* - \Psi_1) \rangle, \quad (8)$$

where $\phi_p^*$ is the decay daughter proton's momentum azimuthal angle in the rest frame of Λ. Specifically, $\phi_p^*$ is the azimuthal angle of $\vec{p}_p^*$ in Eq. 6. The EPD $\Psi_1$ and its resolution $R_{11}$ (Fig. 1) are used to estimate RP. The decay parameters are taken from Ref. [29]:

$$\alpha_\Lambda = 0.732 \pm 0.014, \quad (9)$$

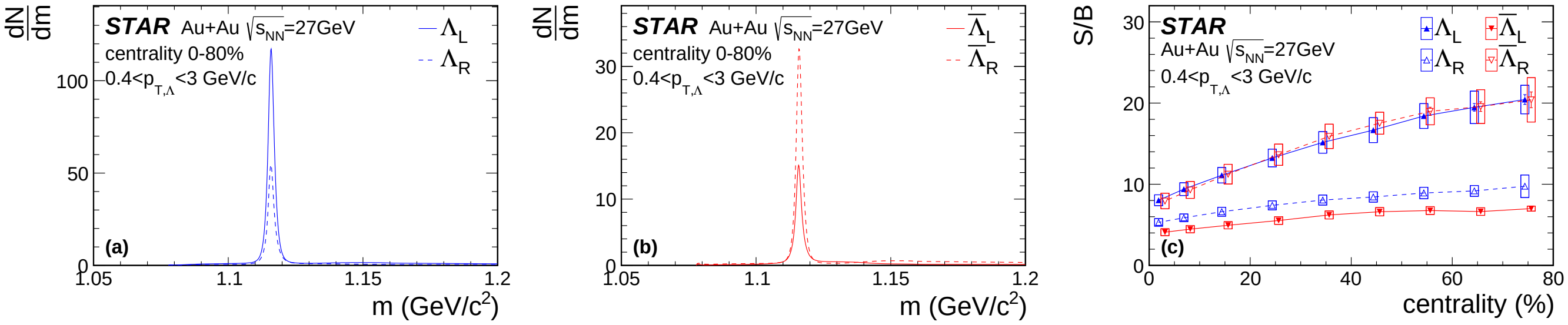

FIG. 4. The reconstructed left- and right-handed $\Lambda$ (a) and $\bar{\Lambda}$ (b) invariant mass spectra in minimum-bias Au+Au collisions at $\sqrt{s_{\rm NN}} = 27$ GeV. (c) Signal to background ratio in the $\Lambda$ mass on-peak region as a function of centrality for each observed handedness. The statistical uncertainty is too small to be visible, while the systematic uncertainty is shown by hollow boxes. The $\Lambda$ data points are shifted slightly to the left along the $x$-axis, while $\bar{\Lambda}$ to the right symmetrically, for better visualization.

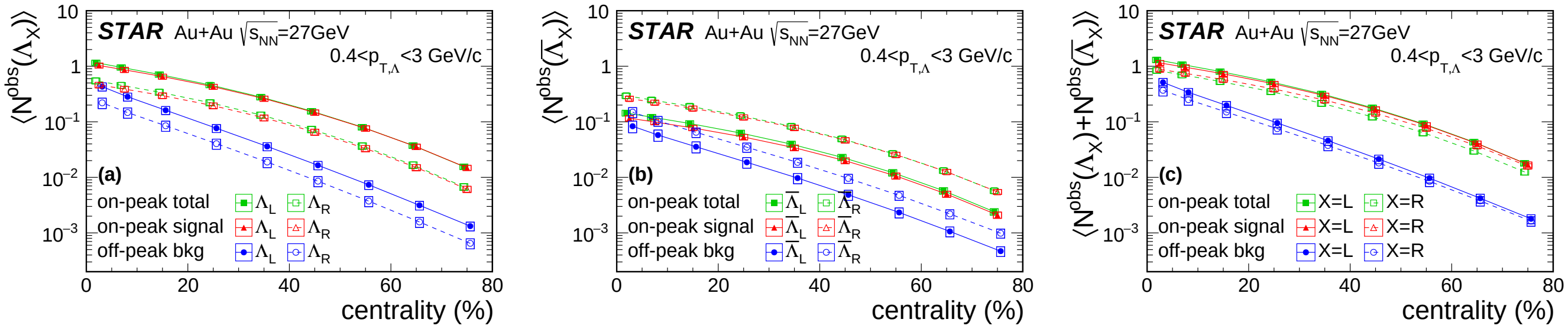

FIG. 5. Observed number of per event of each handedness for (a) $\Lambda$ (b) $\bar{\Lambda}$ and (c) their sum. $\Lambda$ reconstruction inefficiency correction is not included. The statistical uncertainty is too small to be visible, while the systematic uncertainty is shown by hollow boxes. The on-peak total data points are shifted slightly to the left along the $x$-axis, while the off-peak background to the right symmetrically, for better visualization.

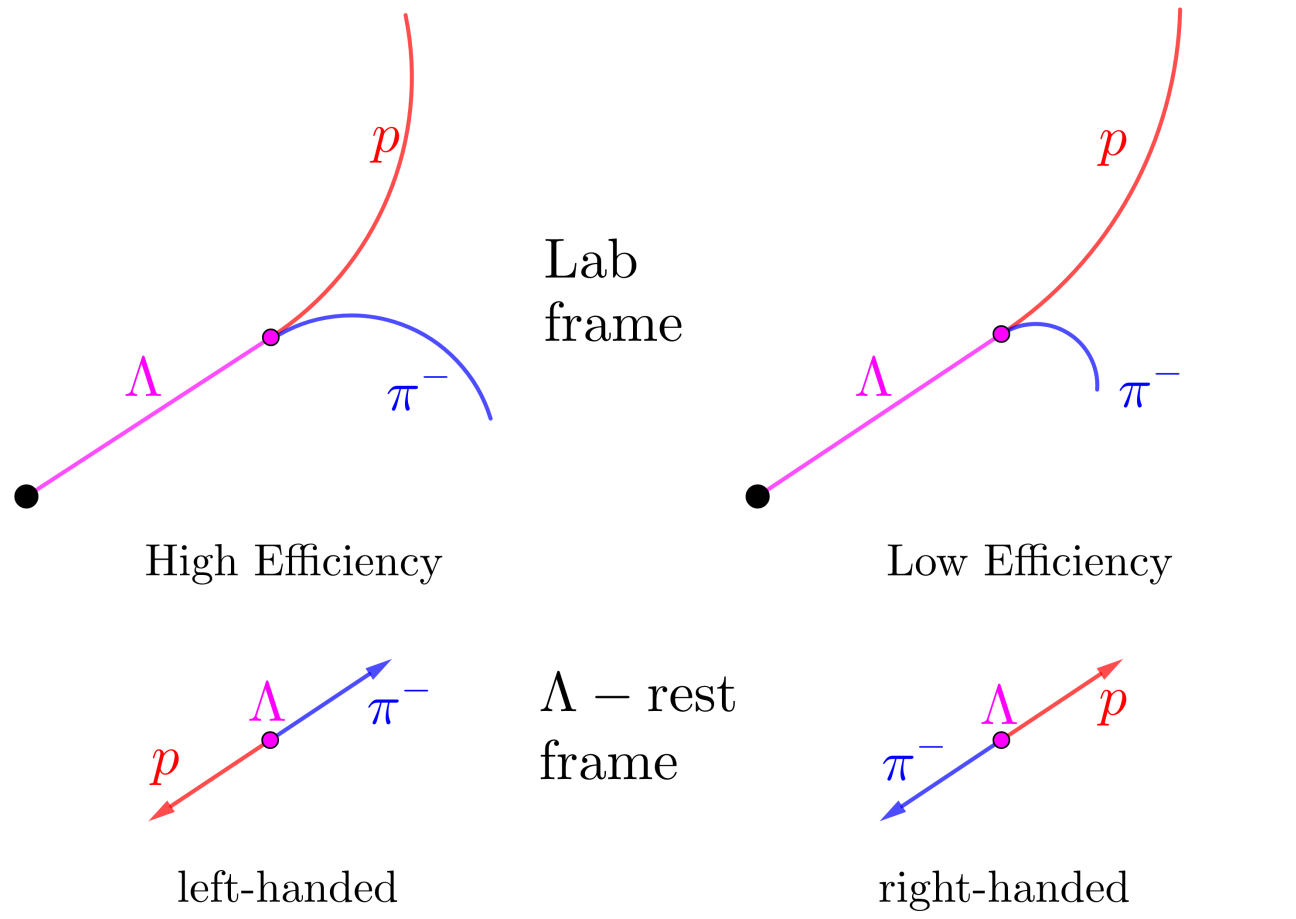

FIG. 6. Schematic diagrams for $\Lambda$ reconstruction inefficiency difference between left- and right-handed $\Lambda$'s. These cartoon are based on Refs. [42, 55]

and $\bar{\Lambda}$ is assumed to have the same value with a minus sign ($\alpha_\Lambda = -\alpha_{\bar{\Lambda}}$).

Before purity correction, the term $\langle \sin(\phi_p^* - \Psi_1)\rangle$ is calculated as a function of centrality in both the $\Lambda$ mass peak region and the off-peak background region (Fig. 2 (b)). Then, the purity correction is [56]

$$\begin{aligned}\langle \sin(\phi_p^* - \Psi_1)\rangle =& \frac{S+B}{S}\langle \sin(\phi_p^* - \Psi_1)\rangle_{\rm peak} \\ &- \frac{B}{S}\langle \sin(\phi_p^* - \Psi_1)\rangle_{\text{off-peak}},\end{aligned} \tag{10}$$

using the signal over background ratio ($S/B$) shown in Fig. 2 (c).

### 2.7. Covariances

The covariances are used to quantify the event-by-event correlations between the parity-odd observables $\Delta n^{\rm obs}$ vs. $\Delta a_1$ (Cov[$\Delta n^{\rm obs}, \Delta a_1$]), and between the parity-even observables $\Delta P$ vs. $\Delta\gamma$ (Cov[$\Delta P, \Delta\gamma$]). The covariance between observables $X$ and $Y$ is given as

$$\begin{aligned}{\rm Cov}[X,Y] =& \langle (X - \langle X\rangle)(Y - \langle Y\rangle)\rangle \\ =& \langle XY\rangle - \langle X\rangle\langle Y\rangle,\end{aligned} \tag{11}$$

where $\langle\cdot\rangle$ means the event average.

To determine Cov[$\Delta P, \Delta\gamma$], the covariances Cov[$P_\Lambda, \Delta\gamma$] and Cov[$P_{\bar{\Lambda}}, \Delta\gamma$] are obtained individually, and their difference is taken. Since the magnetic field acts on $\Lambda$ (and $\bar{\Lambda}$) independently, regardless of whether there is an $\bar{\Lambda}$ (or $\Lambda$) present (or reconstructed) in the same event, this $\mathrm{Cov}[P_\Lambda, \Delta\gamma] - \mathrm{Cov}[P_{\bar{\Lambda}}, \Delta\gamma]$ measurement is equivalent to $\mathrm{Cov}[\Delta P, \Delta\gamma]$.

As discussed in Sec. 2.5, the asymmetry in the detector inefficiency causes an observed non-zero average handedness. However, this detector-induced imbalance does not affect our correlation measurement between $\Delta n^{\mathrm{obs}}$ and $\Delta a_1$, because $\Delta a_1$ measurement has nothing to do with this efficiency asymmetry. The detector-induced imbalance in event-average handedness is automatically canceled in the definition of covariance (Eq. 11).

Likewise, the $\Delta\gamma$ measurement is dominated by physics backgrounds, caused mainly by two-particle correlations coupled with elliptic flow, such as resonance decays [7, 18–22]. The $\Lambda$ polarization, on the other hand, is measured by the decay proton momentum direction in the $\Lambda$-rest frame, with respect to the first-order event plane, and therefore unaffected by the elliptic flow of $\Lambda$. The finite background in $\Delta\gamma$ is thus automatically canceled in the covariance measurement between $\Delta P$ and $\Delta\gamma$.

## 3. SYSTEMATIC STUDY

Systematic uncertainties are assessed by varying analysis selections. The default selection on $V_z$ is $|V_z| < 70$ cm (Sec. 2), and the corresponding variations are $|V_z| < 60$ cm and $|V_z| < 80$ cm. The number of hits for track fitting is required to be $\geq 15$ as the default, and its variations are $\geq 10$ and $\geq 20$. For the CME observables, the primary tracks are selected by requiring DCA $< 1.0$ cm (Sec. 2.3). For this, two variations are examined, DCA $< 0.8$ cm and DCA $< 2.0$ cm. The systematic uncertainty in every reported result in this paper (including individual measurements such as $\Delta a_1$ and covariances) is obtained using the following procedure: one selection criterion is varied at a time, with all other cuts kept at their default values, and thus the deviation is obtained in the result due to changing each selection.

The systematic uncertainty on each result is assigned to be the sum of all those deviations in quadrature, $\sqrt{\sum_i (x_i - x_0)^2/n_i}$. Here $x_0$ denotes the default result, $x_i$ is the result from the $i^{\mathrm{th}}$ systematic variation, and $n_i$ is the number of variations for the given analysis selection. This study does not use Barlow's prescription [57] to subtract statistical fluctuation effects, so our estimation of the systematic uncertainties errs on the conservative side. This is because our measurements are mostly dominated by statistical uncertainties, and a more aggressive assessment of the systematics does not change the qualitative conclusions.

The systematic uncertainties from the various sources are of similar magnitude. The systematic uncertainties are taken to be symmetric, and indicated by open boxes in the figures of this paper. The legends of Figs. 7-10 show the averages over 0-80% and 20-50% centrality ranges with statistical and systematic uncertainties.

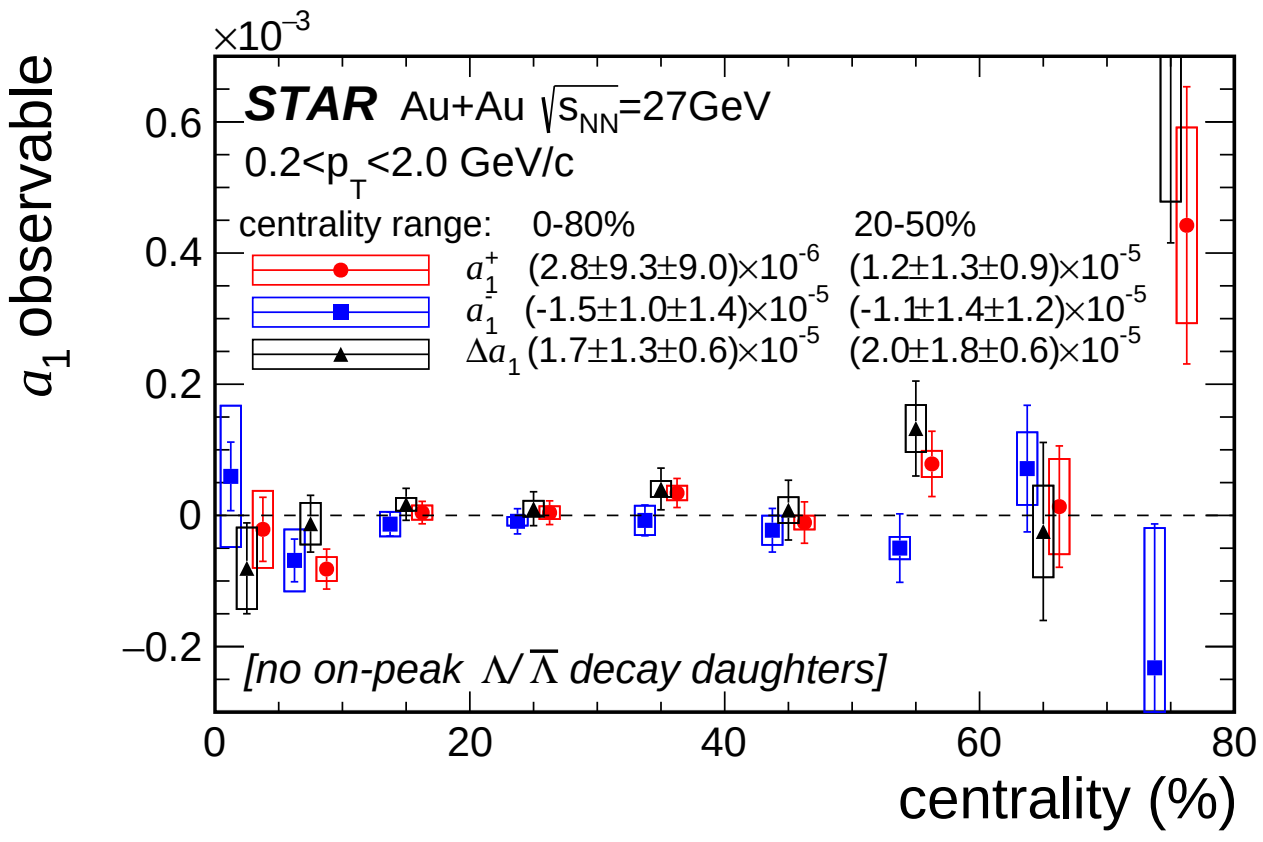

FIG. 7. The $a_1$ observables ($a_1^+$, $a_1^-$, $\Delta a_1$) as functions of centrality in Au+Au collisions at $\sqrt{s_{\mathrm{NN}}} = 27$ GeV. Hadrons used to reconstruct $\Lambda$ or $\bar{\Lambda}$ in the mass peak region are excluded. The statistical uncertainty is shown by error bars, while the systematic uncertainty is shown by hollow boxes. The $a_1^-$ data points are shifted slightly to the left along the $x$-axis, while $a_1^+$ to the right symmetrically, for better visualization.

## 4. RESULTS

The individual measurements of parity-even quantities, including the $\Lambda$ polarizations $P_\Lambda$ ($P_{\bar{\Lambda}}$) and charged-hadron azimuthal correlator $\Delta\gamma$, have been reported by STAR for this dataset [42, 58]. The corresponding measurements from this analysis are in good agreement with those published results.

Figure 7 shows $a_1^+$, $a_1^-$, and $\Delta a_1$ as functions of centrality, calculated from the primary tracks of unidentified charged hadrons, and avoiding the possible self-correlation (Sec. 2.3). All results are found to be consistent with zero, as expected for this parity-odd quantity given that topological charge fluctuations are expected to be random in each event. This is consistent with previous STAR $\Delta a_1$ measurements [6].

The normalized handedness imbalance $\Delta n^{\mathrm{obs}}$ is defined by Eq. 7 event-by-event. The individual measurement of $\Delta n^{\mathrm{obs}}$ is an event average:

$$\langle \Delta n^{\mathrm{obs}} \rangle = \left\langle \frac{N_L^{\mathrm{obs}} - N_R^{\mathrm{obs}}}{\langle N_L^{\mathrm{obs}} + N_R^{\mathrm{obs}} \rangle} \right\rangle = \frac{\langle N_L^{\mathrm{obs}} \rangle - \langle N_R^{\mathrm{obs}} \rangle}{\langle N_L^{\mathrm{obs}} \rangle + \langle N_R^{\mathrm{obs}} \rangle}. \quad (12)$$

It can be directly calculated from $\langle N^{\mathrm{obs}} \rangle$ in Fig. 5. Figure 8 shows $\langle \Delta n^{\mathrm{obs}} \rangle$ for $\Lambda$ (a), $\bar{\Lambda}$ (b), and their sum (c).

As discussed in Sec. 2.5, the $\Lambda$ reconstruction inefficiency detector effect makes $\langle N_L^{\mathrm{obs}}(\Lambda) \rangle \gg \langle N_R^{\mathrm{obs}}(\Lambda) \rangle$ and $\langle N_L^{\mathrm{obs}}(\bar{\Lambda}) \rangle \ll \langle N_R^{\mathrm{obs}}(\bar{\Lambda}) \rangle$, rendering $\Delta n^{\mathrm{obs}}(\Lambda) > 0$ and $\Delta n^{\mathrm{obs}}(\bar{\Lambda}) < 0$. Since more $\Lambda$ hyperons are measured/reconstructed than $\bar{\Lambda}$ due to baryon stopping effect, the inclusive handedness imbalance $\Delta n^{\mathrm{obs}}(\Lambda + \bar{\Lambda}) > 0$. Thus, in Fig. 8, the deviations from zero are solely detector-specific, and not physical. Although the individual measurements of $\Delta n^{\mathrm{obs}}$ (Fig. 8) are influenced by the detector effect of $\Lambda$ reconstruction inefficiency (Sec. 2.5), this automatically cancels out in the correlation covariance (Eq. 11).

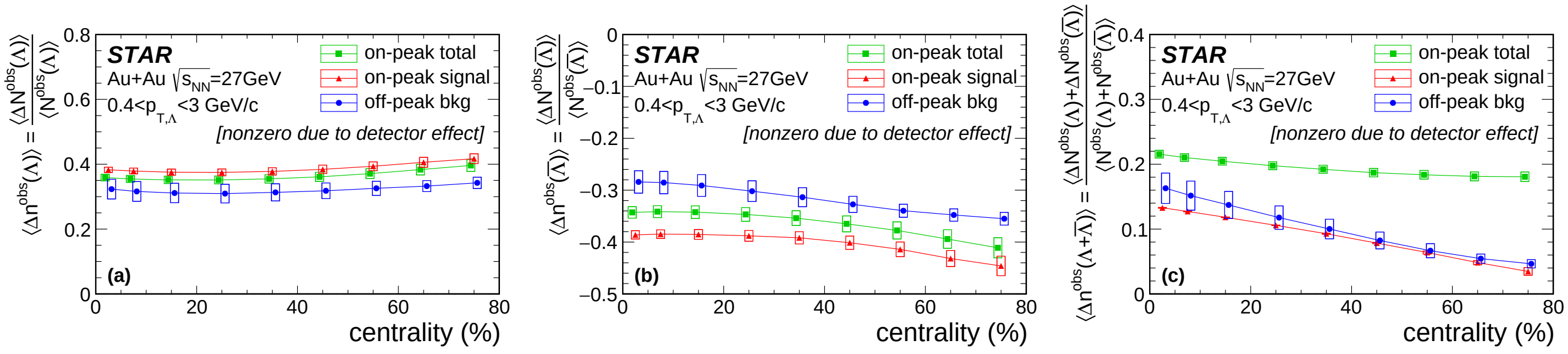

FIG. 8. Observed handedness imbalance $\langle \Delta n^{\rm obs}\rangle$ for $\Lambda$ (left), $\bar{\Lambda}$ (middle), and their sum (right) as functions of centrality in Au+Au collisions at $\sqrt{s_{\rm NN}} = 27$ GeV. If all $\Lambda$'s were reconstructed, the event-average $\langle \Delta n^{\rm obs}\rangle$ would be expected to be zero because this is a parity-odd quantity and the topological charge fluctuations are expected to be random in each event. The non-zero measurements are only due to the detector effect ($\Lambda$ reconstruction inefficiency). The statistical uncertainty is shown by error bars, while the systematic uncertainty is shown by hollow boxes. The on-peak total data points are shifted slightly to the left along the $x$-axis, while the off-peak background to the right symmetrically, for better visualization.

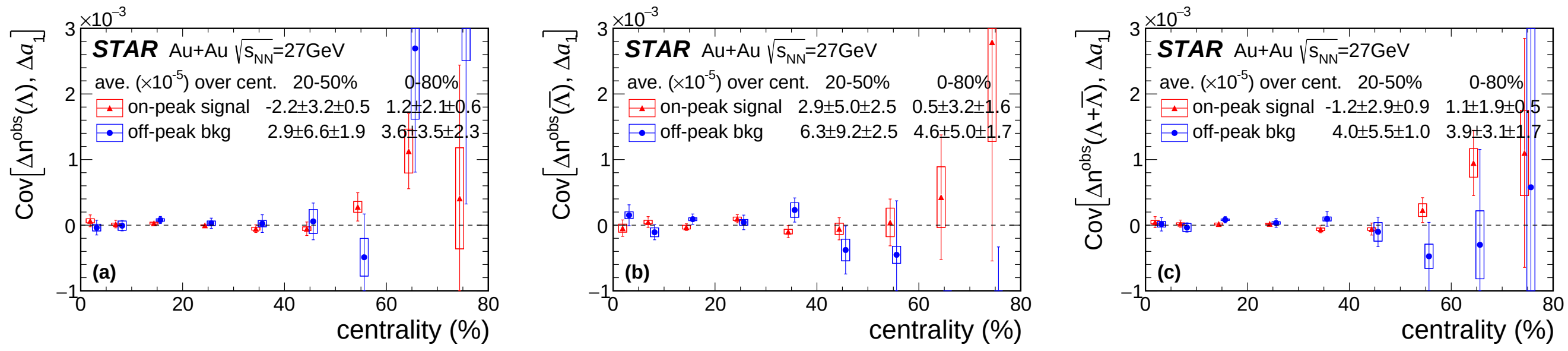

FIG. 9. The covariance between the parity-odd observables $\Delta a_1$ and $\Delta n^{\rm obs}$ for $\Lambda$ (left), $\bar{\Lambda}$ (right), and their sum (right) as functions of centrality in Au+Au collisions at $\sqrt{s_{\rm NN}} = 27$ GeV. Hadrons used to reconstruct $\Lambda$ or $\bar{\Lambda}$ in the mass peak region are excluded from $\Delta a_1$. The statistical uncertainty is shown by error bars, while the systematic uncertainty is shown by hollow boxes. The on-peak signal data points are shifted slightly to the left along the $x$-axis, while the off-peak background to the right symmetrically, for better visualization.

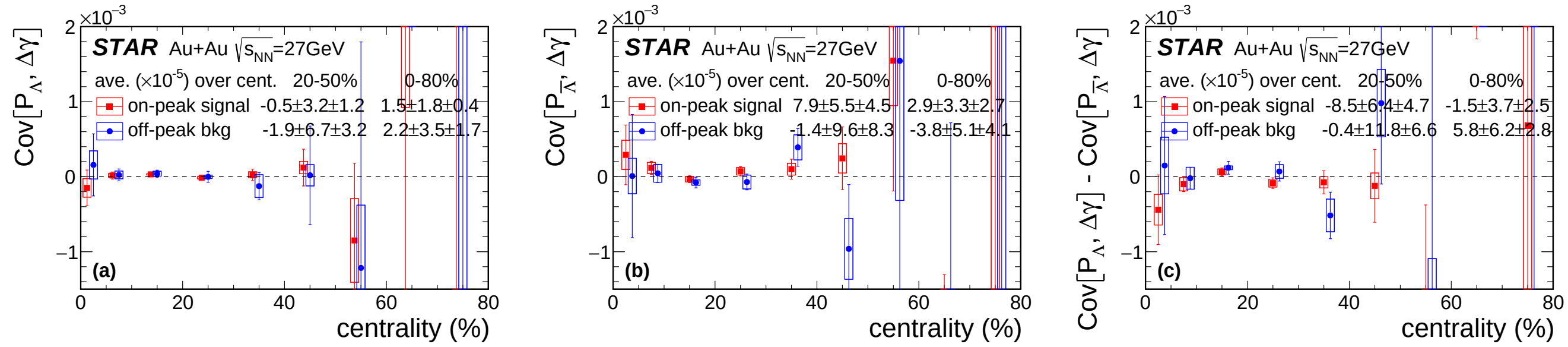

FIG. 10. Covariances between the parity-even observables $P_\Lambda$ and $\Delta\gamma$ (left), between $P_{\bar{\Lambda}}$ and $\Delta\gamma$ (middle), and their difference (right) as functions of centrality in Au+Au collisions at $\sqrt{s_{\rm NN}} = 27$ GeV. Hadrons used to reconstruct $\Lambda$ or $\bar{\Lambda}$ in the mass peak region are excluded from $\Delta\gamma$. The statistical uncertainty is shown by error bars, while the systematic uncertainty is shown by hollow boxes. The on-peak signal data points are shifted slightly to the left along the $x$-axis, while the off-peak background to the right symmetrically, for better visualization.

Figure 9 shows the observed correlation between $\Delta a_1$ and $\Delta n^{\rm obs}$ in each centrality class. Both the signal (using $\Lambda$'s reconstructed in the mass peak) and the background (using off-peak background $\Lambda$'s) covariances are consistent with zero with the current uncertainties.

Figure 10 shows the observed correlation between $\Delta\gamma$ and polarizations (Cov$[P_\Lambda, \Delta\gamma]$, Cov$[P_{\bar{\Lambda}}, \Delta\gamma]$, and their difference) as functions of centrality. With the current statistics, both the signal and background are consistent with zero.

The uncertainties of our measurements are of the order of a few times $10^{-5}$ for both Cov$[\Delta n^{\rm obs}, \Delta a_1]$ and Cov$[\Delta P, \Delta\gamma]$ correlations. Our null results suggest that these correlations are likely smaller than $10^{-4}$. Since the correlation strengths depend on details of the physics underlying the correlations, the implication of our results in terms of the chiral magnetic effect and the magnetic field in heavy-ion collisions requires theoretical input.

## 5. SUMMARY

In conclusion, this paper reports measurements of event-by-event correlations between the observed $\Lambda$ handedness and the charged hadron $\Delta a_1$, and between $\Lambda$ polarizations and charged hadron $\Delta\gamma$, in Au+Au collisions at $\sqrt{s_{\rm NN}} = 27$ GeV using the STAR detector. These correlation observables have been deployed to measure the chiral magnetic effect and the presence of a strong magnetic field in heavy-ion collisions.

Neither of these measurements has yielded a non-zero correlation result within the statistical precision of the present dataset. However, looking toward the future, these correlation measurements should be largely insensitive to the typical physics backgrounds that plague measurements of CME-senstitive observables, and it is possible that such correlation measurements will ultimately offer better sensitivity than individual measurements of these quantities to investigate the chiral magnetic effect.

## ACKNOWLEDGMENTS

We thank the RHIC Operations Group and RCF at BNL, the NERSC Center at LBNL, and the Open Science Grid consortium for providing resources and support. This work was supported in part by the Office of Nuclear Physics within the U.S. DOE Office of Science, the U.S. National Science Foundation, National Natural Science Foundation of China, Chinese Academy of Science, the Ministry of Science and Technology of China and the Chinese Ministry of Education, the Higher Education Sprout Project by Ministry of Education at NCKU, the National Research Foundation of Korea, Czech Science Foundation and Ministry of Education, Youth and Sports of the Czech Republic, Hungarian National Research, Development and Innovation Office, New National Excellency Programme of the Hungarian Ministry of Human Capacities, Department of Atomic Energy and Department of Science and Technology of the Government of India, the National Science Centre and WUT ID-UB of Poland, the Ministry of Science, Education and Sports of the Republic of Croatia, German Bundesministerium für Bildung, Wissenschaft, Forschung and Technologie (BMBF), Helmholtz Association, Ministry of Education, Culture, Sports, Science, and Technology (MEXT) and Japan Society for the Promotion of Science (JSPS).

[1] D. E. Kharzeev, L. D. McLerran, and H. J. Warringa, The Effects of topological charge change in heavy ion collisions: 'Event by event P and CP violation', Nucl.Phys. **A803**, 227 (2008), arXiv:0711.0950 [hep-ph].

[2] D. E. Kharzeev, The Chiral Magnetic Effect and Anomaly-Induced Transport, Prog. Part. Nucl. Phys. **75**, 133 (2014), arXiv:1312.3348 [hep-ph].

[3] D. Kharzeev, R. D. Pisarski, and M. H. G. Tytgat, Possibility of spontaneous parity violation in hot QCD, Phys. Rev. Lett. **81**, 512 (1998), arXiv:hep-ph/9804221.

[4] T. D. Lee, A Theory of Spontaneous T Violation, Phys. Rev. D **8**, 1226 (1973).

[5] T. D. Lee and G. C. Wick, Vacuum Stability and Vacuum Excitation in a Spin 0 Field Theory, Phys. Rev. D **9**, 2291 (1974).

[6] L. Adamczyk *et al.* (STAR), Fluctuations of charge separation perpendicular to the event plane and local parity violation in $\sqrt{s_{NN}} = 200$ GeV Au+Au collisions at the BNL Relativistic Heavy Ion Collider, Phys. Rev. C **88**, 064911 (2013), arXiv:1302.3802 [nucl-ex].

[7] S. A. Voloshin, Parity violation in hot QCD: How to detect it, Phys.Rev. **C70**, 057901 (2004), arXiv:hep-ph/0406311 [hep-ph].

[8] B. Abelev *et al.* (STAR Collaboration), Azimuthal Charged-Particle Correlations and Possible Local Strong Parity Violation, Phys.Rev.Lett. **103**, 251601 (2009), arXiv:0909.1739 [nucl-ex].

[9] B. Abelev *et al.* (STAR Collaboration), Observation of charge-dependent azimuthal correlations and possible local strong parity violation in heavy ion collisions, Phys.Rev. **C81**, 054908 (2010), arXiv:0909.1717 [nucl-ex].

[10] L. Adamczyk *et al.* (STAR Collaboration), Measurement of charge multiplicity asymmetry correlations in high-energy nucleus-nucleus collisions at $\sqrt{s_{NN}} = 200$ GeV, Phys. Rev. **C89**, 044908 (2014), arXiv:1303.0901 [nucl-ex].

[11] L. Adamczyk *et al.* (STAR Collaboration), Beam-energy dependence of charge separation along the magnetic field in Au+Au collisions at RHIC, Phys. Rev. Lett. **113**, 052302 (2014), arXiv:1404.1433 [nucl-ex].

[12] J. Adam *et al.* (STAR Collaboration), Charge-dependent pair correlations relative to a third particle in $p$ + Au and $d$+ Au collisions at RHIC, Phys. Lett. **B798**, 134975 (2019), arXiv:1906.03373 [nucl-ex].

[13] B. Abelev *et al.* (ALICE Collaboration), Charge separation relative to the reaction plane in Pb-Pb collisions at $\sqrt{s_{NN}} = 2.76$ TeV, Phys.Rev.Lett. **110**, 012301 (2013), arXiv:1207.0900 [nucl-ex].

[14] V. Khachatryan *et al.* (CMS Collaboration), Observation of charge-dependent azimuthal correlations in $p$-Pb collisions and its implication for the search for the chiral magnetic effect, Phys. Rev. Lett. **118**, 122301 (2017), arXiv:1610.00263 [nucl-ex].

[15] A. M. Sirunyan *et al.* (CMS Collaboration), Constraints on the chiral magnetic effect using charge-dependent azimuthal correlations in $p$Pb and PbPb collisions at the CERN Large Hadron Collider, Phys. Rev. C **97**, 044912 (2018), arXiv:1708.01602 [nucl-ex].

[16] S. Acharya *et al.* (ALICE Collaboration), Constraining the magnitude of the Chiral Magnetic Effect with Event Shape Engineering in Pb-Pb collisions at $\sqrt{s_{\rm NN}} = 2.76$ TeV, Phys. Lett. **B777**, 151 (2018), arXiv:1709.04723 [nucl-ex].

[17] S. Acharya *et al.* (ALICE Collaboration), Constraining the Chiral Magnetic Effect with charge-dependent azimuthal correlations in Pb-Pb collisions at $\sqrt{s_{\rm NN}} = 2.76$ and 5.02 TeV, JHEP **09**, 160, arXiv:2005.14640 [nucl-ex].

[18] F. Wang, Effects of Cluster Particle Correlations on Local Parity Violation Observables, Phys.Rev. **C81**, 064902 (2010), arXiv:0911.1482 [nucl-ex].

[19] A. Bzdak, V. Koch, and J. Liao, Remarks on possible local parity violation in heavy ion collisions, Phys.Rev. **C81**, 031901 (2010), arXiv:0912.5050 [nucl-th].

[20] S. Schlichting and S. Pratt, Charge conservation at energies available at the BNL Relativistic Heavy Ion Collider and contributions to local parity violation observables, Phys.Rev. **C83**, 014913 (2011), arXiv:1009.4283 [nucl-th].

[21] F. Wang and J. Zhao, Challenges in flow background removal in search for the chiral magnetic effect, Phys. Rev. **C95**, 051901 (2017), arXiv:1608.06610 [nucl-th].

[22] J. Zhao and F. Wang, Experimental searches for the chiral magnetic effect in heavy-ion collisions, Prog. Part. Nucl. Phys. **107**, 200 (2019), arXiv:1906.11413 [nucl-ex].

[23] Y. Feng, J. Zhao, H. Li, H.-j. Xu, and F. Wang, Two- and three-particle nonflow contributions to the chiral magnetic effect measurement by spectator and participant planes in relativistic heavy ion collisions, Phys. Rev. C **105**, 024913 (2022), arXiv:2106.15595 [nucl-ex].

[24] J. Schukraft, A. Timmins, and S. A. Voloshin, Ultra-relativistic nuclear collisions: event shape engineering, Phys. Lett. **B719**, 394 (2013), arXiv:1208.4563 [nucl-ex].

[25] J. Zhao, H. Li, and F. Wang, Isolating the chiral magnetic effect from backgrounds by pair invariant mass, Eur. Phys. J. C **79**, 168 (2019), arXiv:1705.05410 [nucl-ex].

[26] J. Adam *et al.* (STAR Collaboration), Pair invariant mass to isolate background in the search for the chiral magnetic effect in Au+Au collisions at $\sqrt{s_{\rm NN}}$= 200 GeV, (2020), arXiv:2006.05035 [nucl-ex].

[27] H.-j. Xu, J. Zhao, X. Wang, H. Li, Z.-W. Lin, C. Shen, and F. Wang, Varying the chiral magnetic effect relative to flow in a single nucleus-nucleus collision, Chin. Phys. C **42**, 084103 (2018), arXiv:1710.07265 [nucl-th].

[28] L. E. Finch and S. J. Murray, Investigating local parity violation in heavy-ion collisions using $\Lambda$ helicity, Phys. Rev. C **96**, 044911 (2017).

[29] P. A. Zyla *et al.* (Particle Data Group), Review of Particle Physics, PTEP **2020**, 083C01 (2020).

[30] C. Adler *et al.* (STAR Collaboration), Midrapidity Lambda and anti-Lambda production in Au + Au collisions at $\sqrt{s_{NN}} = 130$ GeV, Phys. Rev. Lett. **89**, 092301 (2002), arXiv:nucl-ex/0203016.

[31] F. Du, L. E. Finch, and J. Sandweiss, Observing spontaneous strong CP violation through hyperon helicity correlations, Phys. Rev. C **78**, 044908 (2008).

[32] M. Burkardt and R. L. Jaffe, Polarized q —> Lambda fragmentation functions from e+ e- —> Lambda + X, Phys. Rev. Lett. **70**, 2537 (1993), arXiv:hep-ph/9302232.

[33] M. Gockeler, R. Horsley, D. Pleiter, P. E. L. Rakow, S. Schaefer, A. Schafer, and G. Schierholz (QCDSF), A Lattice study of the spin structure of the Lambda hyperon, Phys. Lett. B **545**, 112 (2002), arXiv:hep-lat/0208017.

[34] J. R. Ellis, D. Kharzeev, and A. Kotzinian, The Proton spin puzzle and lambda polarization in deep inelastic scattering, Z. Phys. C **69**, 467 (1996), arXiv:hep-ph/9506280.

[35] M. Mace, N. Mueller, S. Schlichting, and S. Sharma, Non-equilibrium study of the Chiral Magnetic Effect from real-time simulations with dynamical fermions, Phys. Rev. D **95**, 036023 (2017), arXiv:1612.02477 [hep-lat].

[36] Z.-T. Liang and X.-N. Wang, Globally polarized quark-gluon plasma in non-central A+A collisions, Phys. Rev. Lett. **94**, 102301 (2005), [Erratum: Phys.Rev.Lett. 96, 039901 (2006)], arXiv:nucl-th/0410079.

[37] F. Becattini, L. Csernai, and D. J. Wang, $\Lambda$ polarization in peripheral heavy ion collisions, Phys. Rev. C **88**, 034905 (2013), [Erratum: Phys.Rev.C 93, 069901 (2016)], arXiv:1304.4427 [nucl-th].

[38] B. I. Abelev *et al.* (STAR Collaboration), Global polarization measurement in au+au collisions, Phys. Rev. C **76**, 024915 (2007).

[39] L. Adamczyk *et al.* (STAR Collaboration), Global $\Lambda$ hyperon polarization in nuclear collisions: evidence for the most vortical fluid, Nature **548**, 62 (2017), arXiv:1701.06657 [nucl-ex].

[40] T. Niida (STAR), Global and local polarization of $\Lambda$ hyperons in Au+Au collisions at 200 GeV from STAR, Nucl. Phys. A **982**, 511 (2019), arXiv:1808.10482 [nucl-ex].

[41] M. S. Abdallah *et al.* (STAR), Global $\Lambda$-hyperon polarization in Au+Au collisions at $\sqrt{s_{NN}}$=3 GeV, Phys. Rev. C **104**, L061901 (2021), arXiv:2108.00044 [nucl-ex].

[42] J. Adams, E. Alpatov, M. Lisa, and G. Nigmatkulov (STAR), Global polarization of $\Lambda$ and $\bar{\Lambda}$ hyperons in Au+Au collisions at $\sqrt{s_{\rm NN}} = 19.6$ and 27 GeV, (2023), arXiv:2305.08705 [nucl-ex].

[43] W.-T. Deng and X.-G. Huang, Event-by-event generation of electromagnetic fields in heavy-ion collisions, Phys. Rev. C **85**, 044907 (2012), arXiv:1201.5108 [nucl-th].

[44] A. Bzdak and V. Skokov, Event-by-event fluctuations of magnetic and electric fields in heavy ion collisions, Phys. Lett. B **710**, 171 (2012), arXiv:1111.1949 [hep-ph].

[45] J. Bloczynski, X.-G. Huang, X. Zhang, and J. Liao, Azimuthally fluctuating magnetic field and its impacts on observables in heavy-ion collisions, Phys. Lett. B **718**, 1529 (2013), arXiv:1209.6594 [nucl-th].

[46] H. Li, X.-L. Xia, X.-G. Huang, and H. Z. Huang, Global spin polarization of multistrange hyperons and feed-down effect in heavy-ion collisions, Phys. Lett. B **827**, 136971 (2022), arXiv:2106.09443 [nucl-th].

[47] J. Adams *et al.*, The STAR Event Plane Detector, Nucl. Instrum. Meth. A **968**, 163970 (2020), arXiv:1912.05243 [physics.ins-det].

[48] H. Wieman *et al.* (STAR Collaboration), STAR TPC at RHIC, IEEE Trans. Nucl. Sci. **44**, 671 (1997).

[49] M. Anderson *et al.*, The Star time projection chamber: A Unique tool for studying high multiplicity events at RHIC, Nucl. Instrum. Meth. A **499**, 659 (2003), arXiv:nucl-ex/0301015.

[50] W. J. Llope *et al.*, The TOFp / pVPD time-of-flight system for STAR, Nucl. Instrum. Meth. A **522**, 252 (2004), arXiv:nucl-ex/0308022.

[51] B. I. Abelev *et al.* (STAR), Systematic Measurements of Identified Particle Spectra in $pp, d^+$ Au and Au+Au Collisions from STAR, Phys. Rev. C **79**, 034909 (2009), arXiv:0808.2041 [nucl-ex].

[52] A. M. Poskanzer and S. Voloshin, Methods for analyzing anisotropic flow in relativistic nuclear collisions, Phys.Rev. **C58**, 1671 (1998), arXiv:nucl-ex/9805001 [nucl-ex].

[53] M. Zyzak, *Online selection of short-lived particles on many-core computer architectures in the CBM experiment at FAIR*, Ph.D. thesis (2016), `https://drupal.star.bnl.gov/STAR/system/files/KFParticleTutorial_11.12.2018.pdf`.

[54] S. Gorbunov, *On-line reconstruction algorithms for the CBM and ALICE experiments*, Ph.D. thesis (2013), `http://publikationen.ub.uni-frankfurt.de/frontdoor/index/index/docId/29538`.

[55] J. R. Adams (STAR Collaboration), Differential measurements of $\Lambda$ polarization in Au+Au collisions and a search for the magnetic field by STAR, Nucl. Phys. A **1005**, 121864 (2021).

[56] I. Upsal, *Global polarization of the $\Lambda/\bar{\Lambda}$ system in the STAR BES*, Ph.D. thesis (2018), `https://drupal.star.bnl.gov/STAR/files/UpsalThesisV4.pdf`.

[57] R. Barlow, Systematic errors: Facts and fictions, in *Conference on Advanced Statistical Techniques in Particle Physics* (2002) pp. 134–144, arXiv:hep-ex/0207026.

[58] B. Aboona *et al.* (STAR), Search for the Chiral Magnetic Effect in Au+Au collisions at $\sqrt{s_{\rm NN}} = 27$ GeV with the STAR forward Event Plane Detectors, Phys. Lett. B **839**, 137779 (2023), arXiv:2209.03467 [nucl-ex].